\documentclass{article}
\usepackage{spconf,amsmath,graphicx}
\usepackage{url}
\usepackage{textcomp}
\usepackage{stfloats}
\usepackage{multirow, booktabs}
\usepackage{booktabs}
\usepackage{url}
\usepackage{verbatim}
\usepackage{cite}
\usepackage{soul} 
\usepackage{bm}
\usepackage{hyperref}
\usepackage{multirow}
\usepackage{tablefootnote}
\usepackage{pifont}
\usepackage{xcolor}

\title{Serialized Speech Information Guidance with Overlapped Encoding Separation for Multi-Speaker Automatic Speech Recognition}
\name{Hao~Shi$^{1}$,~Yuan~Gao$^{1}$,~Zhaoheng~Ni$^{2}$,~Tatsuya~Kawahara$^{1}$}
\address{
$^1$Graduate School of Informatics, Kyoto University, Kyoto, Japan\\
$^2$Meta, New York, USA~\thanks{All the experiments were conducted at Kyoto University}
}

\begin{document}
%
\maketitle
\begin{abstract}
Serialized output training (SOT) attracts increasing attention due to its convenience and flexibility for multi-speaker automatic speech recognition (ASR). 
However, it is not easy to train with  attention loss only. 
In this paper, we propose the overlapped encoding separation (EncSep) to fully utilize the benefits of the connectionist temporal classification (CTC) and attention (CTC-Attention) hybrid loss. 
This additional separator is inserted after the encoder to extract the multi-speaker information with CTC losses. 
Furthermore, we propose the serialized speech information guidance SOT (GEncSep) to further utilize the separated encodings. 
The separated streams are concatenated to provide single-speaker information to guide attention during decoding. 
The experimental results on Libri2Mix and Libri3Mix show that the single-speaker encoding can be separated from the overlapped encoding. 
The CTC loss helps to improve the encoder representation under complex scenarios (three-speaker and noisy conditions), which makes the EncSep have a relative improvement of more than 8\% and 6\% on the noisy Libri2Mix and Libri3Mix evaluation sets, respectively. 
GEncSep further improved performance, which was more than 12\% and 9\% relative improvement for the noisy Libri2Mix and Libri3Mix evaluation sets. 
\end{abstract}
\section{Introduction}
\label{sec:intro}
Automatic speech recognition gets impressive performance with the development of deep learning \cite{6423821,10542371,gulati20_interspeech,sun24e_interspeech,sun2024fine,sun2024enhancing}. 
The word error rate (WER) of ASR for single speaker conditions has achieved the level of human transcribers \cite{gulati20_interspeech,6732927}, even when faced with many complex scenarios, such as noise \cite{10094718,chang22g_interspeech, 9689650, SIP-2021-0050, shi2024investigation}. 
However, compared to the additive noise and reverberation, the inference from other speakers, known as the cocktail problem, has a more severe effect on ASR \cite{dang23_interspeech,kanda2019auxiliary,neumann20_interspeech}. 
As a result, ASR performance has dramatic degradation under multi-speaker scenarios \cite{kanda2019auxiliary}.

Recently, extensive research has been conducted on multi-speaker ASR \cite{kanda2019auxiliary,neumann20_interspeech}. 
It is intuitive to decompose the multi-speaker ASR task to speech separation and then recognition \cite{6739096,10448116}. 
The pipeline consists of a speech separation front-end \cite{8369155,10446870} followed by a recognizer \cite{6739096}. 
However, speech separation front-ends, especially the single-channel methods, often bring speech information loss and distortion issues, which harms ASR \cite{6739096, iwamoto2023does}. 
With the development of end-to-end ASR, the ASR back-end already has some abilities to handle multi-speaker conditions \cite{7979557,9054328}. 
Utterance-level permutation invariant training (uPIT) \cite{7979557} is popular for multi-speaker ASR \cite{7952154}. 
During training, all possible permutations of speakers are needed to compute loss, and the smallest loss is used for backpropagation \cite{7952154}. 
In the uPIT-based ASR, the number of output layers constrains the maximum number of speakers \cite{kanda20b_interspeech}, 
and training becomes computationally complex with more speakers \cite{kanda20b_interspeech}.

The serialized output training (SOT) \cite{kanda20b_interspeech} is proposed to solve the above-mentioned drawbacks. 
SOT-based ASR is based on the attention-based encoder-decoder (AED) structure \cite{6af3452a28a04980b2b8f5eb48730d36,chan2015listen}. 
It designs the training label: the overlapped speeches are serialized into a single token sequence according to the speaking start-time of each speaker \cite{kanda20b_interspeech}. 
The problem of the variable speaker number could also be alleviated without the performance degradation compared to uPIT-based ASR \cite{kanda20b_interspeech}. 
Given the success of SOT, more studies improve the model's performance by designing training targets. 
The timestamp of each speaker is inserted as a token to the training label \cite{makishima23_interspeech}. 
Besides, the speaker diarization task also helps the SOT-based ASR \cite{shen2023speaker}.

The connectionist temporal classification (CTC) \cite{graves2006connectionist} enforces the monotonic alignment between speech and label sequences, which helps the attention by solving the misalignment issues \cite{8068205}. 
Thus, the CTC and attention (CTC-Attention) hybrid loss \cite{8068205} is widely adopted for training the ASR systems. 
Although it is commonly used in single-speaker conditions, it is difficult to use it in SOT-based ASR because the serialized training label is difficult to align with the overlapped speech embedding. 
Thus, the SOT-based ASR systems only use attention cross-entropy loss for training \cite{kanda20b_interspeech,makishima23_interspeech,shen2023speaker}.

\begin{figure*}
    \centering
    \includegraphics[width=0.95\linewidth]{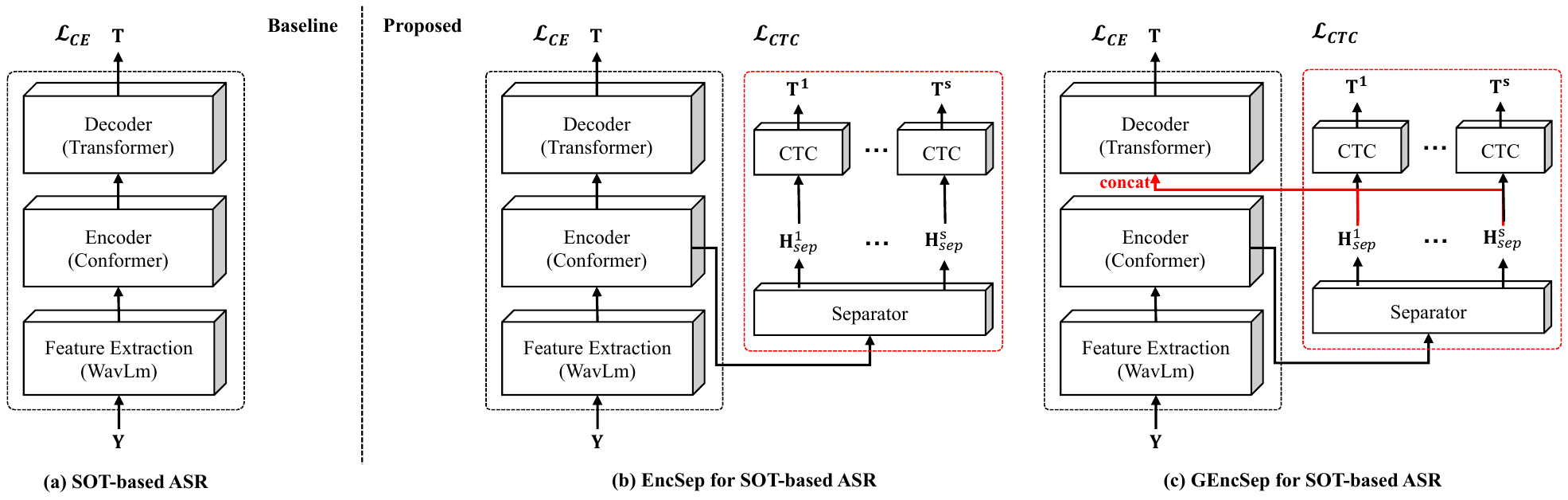}
    \vspace{-10pt}
    \caption{Flowchart of training strategies in different multi-speaker ASR systems.}
    \label{fig:different_training}
\end{figure*}

In this paper, we first propose the overlapped encoding separation (EncSep) to improve the encoding representation in the SOT-based ASR by utilizing the CTC-Attention loss. 
An additional separator is used to get the single speaker's encoding from the encoder's overlapped speech embedding. 
The separated encodings are arranged according to speaking time, and the CTC loss is computed. 
We also propose the single-speaker information guidance SOT (GEncSep) to further utilize the separated encodings, which are concatenated for decoding. 
The attention mechanisms are used to focus on the different single-speaker information from the concatenated embedding.

\section{Related Works}
\label{sec:related_works}
\subsection{Utterance-level Permutation Invariant Training (uPIT) for Multi-speaker ASR}
uPIT-based multi-speaker ASR contains a shared mixed encoder, several speaker-differentiating (SD) encoders, a shared recognition encoder, a shared attention module, and a shared decoder. 
The mixed encoder transforms the overlapped speech feature $\textbf{Y}$ to embedding $\textbf{H}$. 
Then, the extracted mixed embeddings are fed into several SD encoders. 
Each SD encoder only extracts one speaker information $\textbf{H}_{SD}^{s}$, 
$s$ represents the $s$-th speaker. 
The shared recognition encoder transforms the representation of the $s$-th speaker from SD encoders to high-level representations for recognition $\textbf{H}_{Reg}^{s}$. 

Finding a definite relationship between the separated features and multiple speakers is difficult for multi-speaker processing. 
uPIT was proposed to solve this permutation issue by computing the smallest loss with all different estimation-label permutations. 
uPIT is adopted to compute the CTC losses between the encoding of the recognition encoder $\textbf{H}_{Reg}^{s}$ and its corresponding label $\textbf{T}^{s}$. 
An additional linear layer is inserted after the recognition encoder to compute the CTC loss and get the transcription $\textbf{C}^{s}$. 
The number of permutations between the recognition encodings and labels is $S!$, where $S$ represents the speaker number. 
The loss function of CTC with uPIT is as follows:
\begin{equation}
    \mathcal{L}_{\text{CTC-uPIT}} = \arg\min_{\pi \in \mathcal{P}} \sum\nolimits_{s=1}^{S} \text{Loss}_{\text{CTC}}(\textbf{C}^{s}, \textbf{T}^{\pi(s)}),
\end{equation}
where $\mathcal{P}$ represents the set of all permutations of ${1, ..., S}$. 
$\pi(s)$ represents the $s$-th element in a permutation $\pi$, and $\textbf{T}$ is the set of transcription labels for $S$ speakers. 
The recognition encodings $\textbf{H}_{Reg}$ are finally fed into the shared attention and the shared decoder to get the attention-based outputs $\textbf{H}_{Att}^{s}$. 
The loss of the decoder is cross-entropy (CE) with the same permutation as that of the CTC-uPIT: 
\begin{equation}
    \mathcal{L}_{\text{CE}} = \sum\nolimits_{s=1}^{S} \text{Loss}_\text{CE}(\textbf{H}_{Att}^{s}, \textbf{T}^{\hat{\pi}(s)})
\end{equation}
where $\hat{\pi}(s)$ represents the permutation obtained from the CTC-uPIT. 
The final training loss for uPIT-based ASR is CTC-Attention Hybrid loss: 
\begin{equation}
    \mathcal{L}_{\text{uPIT}} = \lambda \mathcal{L}_{\text{CTC-uPIT}} + (1 - \lambda) \mathcal{L}_{\text{CE}}.
\end{equation}
$\lambda$ is the hyperparameter to control the two losses.

\subsection{Serialized Output Training (SOT) for Multi-speaker ASR}
SOT is based on the attention-based encoder-decoder (AED) structure to solve the multi-speaker ASR, which is shown in Fig.~\ref{fig:different_training}--(a). 
It contains an encoder, an attention module, and a decoder. 
All three components cooperate during the decoding to output transcriptions of different speakers according to their start-time of speaking. 
The encoder transforms the overlapped speech feature $\textbf{Y}$ to embedding $\textbf{H}$. 
Different from the uPIT-based ASR, the overlapped speech embedding is directly fed into the attention and decoder to get the transcriptions without any implicit or explicit separation. 
Training targets help the model achieve this ability. 
It innovatively arranges the transcriptions of different speakers according to their start-time of speaking to form a new transcription. 
A special symbol $\langle sc \rangle$ is inserted between different speakers to represent the speaker change. 
For example, for a two-speaker case, the label will be given as $\textbf{T} = \{t_1^{1}, ..., t_1^{N^{1}}, \langle sc \rangle, t_2^{1}, ..., t_2^{N^{2}} \}$, where $t_1$ and $t_2$ represent the transcriptions of the $1$-th and $2$-nd speaker, respectively. 
The $N^{1}$ and $N^{2}$ represent the lengths of the transcriptions. 
Based on this training target, the attention mechanism is able to focus on the required information in encoding the overlapped speech and decoding the transcriptions of multiple speakers $\textbf{C}$ according to their speaking time. 
The loss function of SOT-based ASR is simple as follows: 
\begin{equation}
    \mathcal{L}_{\text{SOT}} = \text{Loss}_\text{CE}(\textbf{C}, \textbf{T}),
    \label{eq:sot_ce}
\end{equation}
Because the embedding obtained by the encoder does not perform any separation, it is hard to align the speech embedding with the label sequence by CTC. 
As a result, only CE loss is used when training the SOT-based ASR system.

\subsection{Self-Supervised Learning-based Feature Extraction}
Feature extraction with self-supervised learning (SSL) shows powerful performance improvement for ASR \cite{chang22g_interspeech}. 
The outputs of the SSL module replace the traditional features like Mel-Frequency Cepstral Coefficients (MFCC) and Filter Bank (F-Bank). 
The SSL-based feature extraction module \cite{shi2023emotion,he2024mf,tian2023semi} typically consists of convolution layers and transformer, e.g., WavLM \cite{9814838}, HuBERT \cite{10096630}. 
With massive pretraining data, the SSL models already contained the capability to convert the speech waveform into the acoustic hidden units. 
Moreover, some SSL models \cite{9814838, 10096630} demonstrate robustness against noise or other inference speakers by incorporating noisy simulation. 
The other parts of the ASR system are the same as the other E2E ASR systems.

\section{Proposed Method}
\label{sec:proposed_method}
To improve the encoder's representation, we first propose the overlapped encoding separation (EncSep) to utilize the CTC-Attention hybrid loss in the SOT-based ASR. 
Then, we propose the single-speaker information guidance SOT (GEncSep) to utilize the separated embeddings.

\subsection{Overlapped Encoding Separation (EncSep)}
In the SOT-based ASR, the encoder transforms the overlapped speech feature $\textbf{Y}$ to embedding $\textbf{H}$. 
The embedding $\textbf{H}$ here is overlapped by speakers. 
Thus, some module for speech separation is introduced to convert the overlapped encoding $\textbf{H}$ into separated single-speaker encodings $\textbf{H}_{sep}$: 
\begin{equation}
    \textbf{H}_{sep} = \text{Separator}(\textbf{H}).
\end{equation}
The Long Short-Term Memory (LSTM) \cite{Graves2012} is adopted as the separator in this work: 
\begin{equation}
    \begin{split}
        & \hat{\textbf{H}} = \text{LayerNorm}(\text{LSTM}(\textbf{H})), \\
        & \textbf{H}_{sep}^{s} = \text{ReLU}(\text{Linear}^{s}(\hat{\textbf{H}})), 
    \end{split}
\end{equation}
Several linear layers are used to generate the single-speaker information $\textbf{H}_{sep}^{s}$. 
Linear layers correspond to a speaker according to the start time of speaking in a serialized manner. 
Then, the CTC loss is computed as follows: 
\begin{equation}
\begin{split}
    \mathcal{L}_{\textbf{CTC-EncSep}} & = \sum\nolimits_{s=1}^{S}\text{Loss}_{\text{CTC}}(\mathbf{C}^{s}, \mathbf{T}^{s}) \\
    & = \sum\nolimits_{s=1}^{S}\text{Loss}_{\text{CTC}}(\text{Linear}^{s}(\mathbf{H}_{sep}^{s}), \mathbf{T}^{s})
\end{split}
\end{equation}
where $\textbf{T}^{s}$ represents the $s$-th transcription arranged in the serialized manner.

The attention-based CE loss, same as Eqn.~(\ref{eq:sot_ce}), is also used for training. 
The final loss function for training EncSep is as follows:
\begin{equation}
    \mathcal{L}_{\text{EncSep}} = \gamma \mathcal{L}_{\text{CTC-EncSep}} + (1 - \gamma)\mathcal{L}_{\text{SOT}},
    \label{eq:encsep}
\end{equation}
where the $\gamma$ is the hyperparameter to balance the two losses. 
The flowchart of the proposed EncSep training strategy is shown in Fig.~\ref{fig:different_training}--(b). 
The separator is only used to introduce CTC information during training. 
When decoding, EncSep maintains the same structure as the SOT-based method (Fig.~\ref{fig:different_training}--(a)). 
Thus, EncSep does not increase any computational cost during decoding from the SOT-based method. 

\subsection{Single-speaker Information Guidance SOT (GEncSep)}
EncSep only uses the separated embeddings $\textbf{H}_{sep}$ to introduce the CTC loss into the encoder. 
To further utilize the separated embeddings $\textbf{H}_{sep}$, we also propose the single-speaker information guidance SOT (GEncSep). 
Compared with the EncSep, the GEncSep utilizes the separated embeddings $\textbf{H}_{sep}$ from the separator, which is shown in Fig.~\ref{fig:different_training}--(c).

GEncSep contains an encoder, which transforms the overlapped speech feature $\textbf{Y}$ to embedding $\textbf{H}$. 
The separator separates the overlapped embedding $\textbf{H}$ into single-speaker embeddings $\textbf{H}_{sep}^{1}$, ..., $\textbf{H}_{sep}^{S}$. 
Then, the separated embeddings are concatenated over time dimension as: 
\begin{equation}
    \textbf{H}_{con} = \text{Concat}(\textbf{H}_{sep}^{1}, ..., \textbf{H}_{sep}^{S})
\end{equation}
The attention mechanism is used to compute the attention weights with single-speaker information: 
\begin{equation}
    \textbf{a}_{con}^{n} = \text{Attention} \left( \textbf{H}_{con}, \textbf{d}^{n-1} \right).
\end{equation}
$\textbf{a}_{con}^{n}$ represents the context vector obtained with the attention mechanism from the concatenated embedding $\textbf{H}_{con}$. 
$\textbf{d}^{n-1}$ is the hidden state of the decoder. 
The decoder decodes from the attention feature and also the previously predicted tokens: 
\begin{equation}
    \textbf{c}^{n} = \text{Decoder} \left( \textbf{H}_{con}, \textbf{a}_{con}^{n}, \textbf{c}^{1:n-1} \right). 
\end{equation}
The corresponding output $\textbf{C}$ is generated iteratively by the decoder. 
The training loss function of GEncSep is the same as Eqn.~(\ref{eq:encsep}).

\section{Experiments}
\label{sec:experiments}
\subsection{Datasets and Experimental Settings}
We used the LibriMix dataset \cite{cosentino2020librimix} to evaluate the model performance. 
It used the train-clean-100, train-clean-360, dev-clean, and test-clean subsets from the LibriSpeech dataset \cite{7178964} as the clean speech. 
The noise samples were taken from WHAM! dataset \cite{Wichern2019WHAM}. 
We used the official scripts\footnote{\url{https://github.com/JorisCos/LibriMix}} to synthesize Libri2Mix and Libri3Mix. 
We used the offset file to make different speaking start times for multiple speakers. 
The two-speaker offset files followed the official ESPnet setting
\footnote{\url{https://github.com/espnet/espnet/tree/master/egs2/librimix/sot_asr1}}
, and the three-speaker offset files were created by ourself.

Conformer \cite{gulati20_interspeech} was used as the ASR back-end. 
Except for the proposed modules, the other modules followed the official ESPnet settings\footnote{\url{https://github.com/espnet/espnet/blob/master/egs2/librimix/sot_asr1/conf/tuning/train_sot_asr_conformer_wavlm.yaml}}. 
The baselines and EncSep used the original Transformer layer. 
For GEncSep, the structure of the decoder layer is shown in Fig.~\ref{fig:different_training}--(c). 
A character-based vocabulary was used with the size 32, including a $\langle sc \rangle$ for speaker change symbol. 
We used the WavLM-Large \cite{9814838} for feature extraction, which has been shown effective for noise-robust ASR \cite{chang2022endtoend}. 
During training, all parameters of the WavLM-Large were frozen. 
In the proposed method, the input and output dimensions of the separator were 256 and 512; the number of LSTM layers were 2. 
The dimensions of linear layers were 512 and 256 for input and output, respectively. 
We also tried bidirectional LSTM layers. 
For bidirectional LSTM, the dimensions of linear layers were 1024 and 256 for input and output, respectively. 

For baselines, the ``\textbf{SOT}'' followed the ESPnet official setting; only the attention CE loss was used with Eqn.~(\ref{eq:sot_ce}). 
``\textbf{SOT-H}'' was selected as another baseline. 
Different from the ``\textbf{SOT}'', it used the CTC-Attention hybrid loss instead of only Attention loss with the serialized label to train the model. 
Its neural network structure followed the ESPnet setting.

The training epochs were 60, and the final evaluation model was averaged over ten checkpoints, according to the loss in the validation set. 
It should be emphasized that pretraining is unnecessary for the two-mixed condition, but a few epochs of original SOT training are needed for the three-mixed condition. 
The all parameters of GEncSep was pretrained with EncSep.

\begin{table}[h]
    \renewcommand{\arraystretch}{1.2}
    \caption{Experimental Results on \textbf{Noisy} Libri2Mix and Libri3Mix. ``Bi'' represents the Bi-directional.}
    \vspace{5pt}
    \centering
    \begin{tabular}{c|p{1.5cm}|c|c|c|c|c}
    \toprule[1pt]
    \multirow{2}{*}{\textbf{Exp.}}
    &\multirow{2}{*}{\textbf{Systems}}
    & \multirow{2}{*}{\textbf{Bi}}
    & \multicolumn{2}{c|}{\textbf{Libri2Mix}} & \multicolumn{2}{c}{\textbf{Libri3Mix}} \\
    \cline{4-7}
    &  & & \multicolumn{1}{p{0.7cm}|}{\centering 
     \textbf{Dev}}           & \multicolumn{1}{p{0.7cm}|}{\centering 
     \textbf{Eval}}           & \multicolumn{1}{p{0.7cm}|}{\centering 
     \textbf{Dev}}           & \multicolumn{1}{p{0.7cm}}{\centering 
     \textbf{Eval}}      \\
    \midrule[0.7pt]
    1 & SOT        &   -     & 19.4 & 17.1 & 30.5 & 28.2 \\
    2 & SOT-H      &    -    & 24.6 & 22.0 & -    & -    \\
    \midrule[0.7pt]
    3 & EncSep & \ding{55}  & 18.4 & 15.9 & 28.5 & 26.5 \\
    4 & EncSep & \ding{51}  & 18.0 & 15.7 & 28.6 & 26.4 \\
    \midrule[0.7pt]
    5 & GEncSep & \ding{55} & 18.3	&15.3	&28.8	&25.7 \\
    6 & GEncSep & \ding{51} & 17.2	&15.0	&28.0	&25.9 \\
    \bottomrule[1pt]
    \end{tabular}
    \label{table:noisy_libri}
\end{table}

\begin{table}[h]
    \renewcommand{\arraystretch}{1.2}
    \caption{Experimental Results on \textbf{Clean} Libri2Mix and Libri3Mix. ``Bi'' represents the Bi-directional.}
    \vspace{5pt}
    \centering
    \begin{tabular}{c|p{1.5cm}|c|c|c|c|c}
    \toprule[1pt]
    \multirow{2}{*}{\textbf{Exp.}}
    & \multirow{2}{*}{\textbf{Systems}}
    & \multirow{2}{*}{\textbf{Bi}}
    & \multicolumn{2}{c|}{\textbf{Libri2Mix}} & \multicolumn{2}{c}{\textbf{Libri3Mix}} \\
    \cline{4-7}
     & &  & \multicolumn{1}{p{0.7cm}|}{\centering 
     \textbf{Dev}}           & \multicolumn{1}{p{0.7cm}|}{\centering 
     \textbf{Eval}}           & \multicolumn{1}{p{0.7cm}|}{\centering 
     \textbf{Dev}}           & \multicolumn{1}{p{0.7cm}}{\centering 
     \textbf{Eval}}      \\
    \midrule[0.7pt]
    7 & SOT          &  -    & 6.8  & 7.0  & 15.0 & 14.7 \\
    8 & SOT-H        &   -   & 10.4 & 10.3 & -    & -    \\
    \midrule[0.7pt]
    9 & EncSep & \ding{55}  & 7.0 & 7.3 & 13.9 & 13.4 \\
    10 & EncSep & \ding{51}  & 7.0 &	7.2 & 13.9 & 13.5  \\
    \midrule[0.7pt]
    11 & GEncSep & \ding{55} & 6.7 & 6.8 & 13.0 & 14.3 \\
    12 & GEncSep & \ding{51} & 6.4 & 6.6 & 13.3	& 13.1 \\
    \bottomrule[1pt]
    \end{tabular}
    \label{table:clean_libri}
\end{table}

\begin{table*}[h]
    \renewcommand{\arraystretch}{1.2}
    \caption{Comparison of ASR systems on Libri2Mix and Libri3Mix. ``SSL'' represents the self-supervised learning.}
    \vspace{5pt}
    \centering
    \begin{tabular}{p{4.5cm}|c|p{1.5cm}|p{1.cm}|p{1.cm}|p{1.cm}|p{1.cm}}
    \toprule[1pt]
    \multirow{2}{*}{\textbf{Systems}} & \multirow{2}{*}{\textbf{w/ Front-end}} & \multicolumn{1}{c|}{\multirow{2}{*}{\textbf{w/ SSL}}} & \multicolumn{2}{c|}{\textbf{Libri2Mix}} & \multicolumn{2}{c}{\textbf{Libri3Mix}} \\
    \cline{4-7}
    & & & \multicolumn{1}{c|}{\textbf{Dev}}          & \multicolumn{1}{c|}{\textbf{Eval}}           & \multicolumn{1}{c|}{\textbf{Dev}}           & \multicolumn{1}{c}{\textbf{Eval}}           \\
    
    \midrule[0.7pt]
    \multicolumn{7}{c}{\textbf{Noisy}} \\
    \hline
    PIT-Conformer\tablefootnote{\url{https://github.com/espnet/espnet/tree/master/egs2/librimix/asr1}} & \ding{55} & \multicolumn{1}{c|}{\ding{55}} & \multicolumn{1}{c|}{23.7} &\multicolumn{1}{c|}{23.5} & \multicolumn{1}{c|}{-} & \multicolumn{1}{c}{-} \\
    Conditional-Conformer \cite{guo21_interspeech} & \ding{55}& \multicolumn{1}{c|}{\ding{55}} & \multicolumn{1}{c|}{24.5} & \multicolumn{1}{c|}{24.9} & \multicolumn{1}{c|}{-} & \multicolumn{1}{c}{-} \\
    WavLM Base \cite{zhangweakly} & \ding{55}& \multicolumn{1}{c|}{\ding{51}} & \multicolumn{1}{c|}{-} & \multicolumn{1}{c|}{27.5} & \multicolumn{1}{c|}{-} & \multicolumn{1}{c}{-} \\
    TS-HuBERT \cite{zhangweakly} & \ding{55} & \multicolumn{1}{c|}{\ding{51}} & \multicolumn{1}{c|}{-} & \multicolumn{1}{c|}{24.8} & \multicolumn{1}{c|}{-} & \multicolumn{1}{c}{-} \\
    SOT-Conformer\tablefootnote{\url{https://github.com/espnet/espnet/tree/master/egs2/librimix/sot_asr1}} & \ding{55} & \multicolumn{1}{c|}{\ding{51}} & \multicolumn{1}{c|}{19.4} & \multicolumn{1}{c|}{17.1} & \multicolumn{1}{c|}{30.5} & \multicolumn{1}{c}{28.2} \\
    TSE-Whisper \cite{10389752} & \ding{51} & \multicolumn{1}{c|}{\ding{55}} & \multicolumn{1}{c|}{-} & \multicolumn{1}{c|}{12.0} & \multicolumn{1}{c|}{-} & \multicolumn{1}{c}{-} \\
    GEncSep (this study) & \ding{55} & \multicolumn{1}{c|}{\ding{51}} & \multicolumn{1}{c|}{17.2} & \multicolumn{1}{c|}{15.0} & \multicolumn{1}{c|}{28.0} & \multicolumn{1}{c}{25.9} \\
    \hline
    \multicolumn{7}{c}{\textbf{Clean}} \\
    \hline
    W2V-baseline \cite{10095295} & \ding{55}& \multicolumn{1}{c|}{\ding{51} }&  \multicolumn{1}{c|}{11.6} & \multicolumn{1}{c|}{12.3}  & \multicolumn{1}{c|}{-} & \multicolumn{1}{c}{-} \\
    W2V-Sidecar-ft \cite{10095295}& \ding{55} & \multicolumn{1}{c|}{\ding{51}}& \multicolumn{1}{c|}{7.7} & \multicolumn{1}{c|}{8.1} & \multicolumn{1}{c|}{-} & \multicolumn{1}{c}{-} \\
    WavLM Base+ \cite{10097139} & \ding{55} & \multicolumn{1}{c|}{\ding{51}}& \multicolumn{1}{c|}{-} & \multicolumn{1}{c|}{8.3} & \multicolumn{1}{c|}{-} & \multicolumn{1}{c}{-} \\
    WavLM Base+ \cite{10097139} & \ding{51} & \multicolumn{1}{c|}{\ding{51}}& \multicolumn{1}{c|}{-} & \multicolumn{1}{c|}{7.6} & \multicolumn{1}{c|}{-} & \multicolumn{1}{c}{-} \\
    TSE-CLN \cite{10097139} & \ding{51} & \multicolumn{1}{c|}{\ding{51}}&\multicolumn{1}{c|}{7.1} & \multicolumn{1}{c|}{7.6} & \multicolumn{1}{c|}{-} &\multicolumn{1}{c}{-} \\
    C-HuBERT LARGE \cite{10096630} & \ding{55} & \multicolumn{1}{c|}{\ding{51}}&\multicolumn{1}{c|}{6.6} &\multicolumn{1}{c|}{7.8} & \multicolumn{1}{c|}{-} &\multicolumn{1}{c}{-} \\
    SOT-Conformer & \ding{55} & \multicolumn{1}{c|}{\ding{51}}& \multicolumn{1}{c|}{6.8} & \multicolumn{1}{c|}{7.0} & \multicolumn{1}{c|}{15.0} & \multicolumn{1}{c}{14.7} \\
    GEncSep (this study) & \ding{55} & \multicolumn{1}{c|}{\ding{51}} & \multicolumn{1}{c|}{6.4} & \multicolumn{1}{c|}{6.6} & \multicolumn{1}{c|}{13.3} & \multicolumn{1}{c}{13.1} \\
    \bottomrule[1pt]
    \end{tabular}
    \label{table:noisy_litra}
\end{table*}

\subsection{Experimental Results}
Table~\ref{table:noisy_libri} shows the experimental results on noisy Libri2Mix and Libri3Mix sets. 
Compared with ``SOT'', ``SOT-H'' had a significant performance degradation on both the development and evaluation sets of Libri2Mix with the CTC-Attention hybrid loss, which confirms that the encoder's implicit separation performed poorly (comparison between Exp.~1 and Exp.~2). 
With the separator, the proposed ``EncSep'' significantly improved from ``SOT'' (p-value $<$ 0.01) for both the development and evaluation sets of Libri2Mix and Libri3Mix (comparison between Exp.~1 and Exp.~3). 
It was shown that the CTC losses with separator benefited the encoder representation since ``EncSep'' has the same structure as ``SOT'' during decoding. 
The bi-directional separator did not bring further improvement (comparison between Exp.~3 and Exp.~4). 
For ``GEncSep'', the experimental results suggested that providing the separated embedding helped the decoding, significantly improving the performance on both the development and evaluation sets of Libri2Mix and Lib3Mix compared with ``SOT'' (p-value $<$ 0.01, comparison between Exp.~1 and Exp.~5). 
Furthermore, the bi-directional separator showed significant improvement (p-value $<$ 0.01, comparison between Exp.~5 and Exp.~6) on development sets of Libri2Mix and Libri3Mix, which were used to select the evaluation model.

Table~\ref{table:clean_libri} shows the experimental performance on clean Libri2Mix and Libri3Mix sets. 
Similar to the noisy conditions, ``SOT-H'' had a significant performance degradation from ``SOT'' (comparison between Exp.~7 and Exp.~8). 
The ``EncSep'' did not show any improvement on Libri2Mix (comparison between Exp.~7 and Exp.~9). 
Compared with the noisy two-speaker condition, the ``SOT'' already had strong abilities for encoding the clean two-speaker features. 
CTC helped the encoder improve its encoding capabilities in more complex scenarios. 
The proposed ``EncSep'' still significantly improved the three-speaker conditions (p-value $<$ 0.01, comparison between Exp.~7 and Exp.~9). 
The experimental results for ``GEncSep'' suggested that providing the separated embedding also helped the decoding under clean conditions, especially for three-speaker conditions (all the Libri2Mix and Libri3Mix evaluation sets are significantly improved from ``SOT'', p-value $<$ 0.01, comparison between Exp.~7 and Exp.~12).

We compared several noise robust end-to-end ASR systems in the literature (after 2020) in Table~\ref{table:noisy_litra}. 
It should be emphasized that although the proposed method showed compelling performance in the end-to-end style for multi-speaker ASR, there is still a gap compared to the systems with pipeline systems (speech separation front-end with recognizer, TSE-Whisper) under noisy two-speaker conditions. 
Another possible reason for such a gap is that whisper's training data is massive, while other ASR back-ends only use the noisy Libri2Mix training set. 
For clean sets, the advantages of the front-end did not perform as well as the ``SOT-Conformer''. 
Compared with ``SOT-Conformer'', the proposed GEncSep had significant performance improvement (p-value $<$ 0.05). 
Compared with other methods, the SOT-based multi-speaker ASR system has advantages for more speakers.

\section{Conclusions and Future Works}
\vspace{-5pt}
In this paper, we focused on improving the serialized output training (SOT) for ASR. 
We first proposed the overlapped encoding separation (EncSep) to fully utilize the benefits of the CTC and attention hybrid loss. 
The experimental results on Libri2Mix and Libri3Mix datasets show that the single-speaker encoding can be separated from the overlapped encoding. 
The CTC losses with separator benefit the encoder representation. 
Then, we proposed the serialized speech information guidance SOT (GEncSep) to further utilize the separated information. 
GEncSep further improved performance with the separated embeddings for decoding. 
Compared with the clean Libri2Mix and Libri3Mix, the proposed method has more significant advantages over the more complex, noisy Libri2Mix and Libri3Mix. 
As a result, the proposed GEncSep had more than 12\% and 9\% relative improvement for the noisy Libri2Mix and Libri3Mix evaluation sets compared to the original SOT. 
In the future, we will make the system use different information sources by fusing the overlapped and separated embeddings.

\clearpage

\bibliographystyle{IEEEbib}
\bibliography{refs}

\end{document}